\documentstyle[12pt,epsf]{article}

\begin{document}

\renewcommand{\thepage}{}

\begin{titlepage}

\title{
\hfill
\parbox{5cm}
{}\\
\vspace{5mm}
CP and T violation test in neutrino oscillation
}
\author{
Joe Sato\thanks{e-mail address:
 {\tt joe@icrhp3.icrr.u-tokyo.ac.jp}}\\
 {\it Institute for Cosmic Ray Research,
   University of Tokyo, Midori-cho,}\\
      {\it Tanashi, Tokyo 188, Japan}}
\date{\today}

\maketitle

\begin{abstract}
I examine how large violation of CP and T is allowed
in long base line neutrino experiments.
When we attribute both the atmospheric neutrino anomaly
and the solar neutrino deficit to
neutrino oscillation we may have a sizable T violation
effect proportional to the ratio of two mass differences;
it is difficult to see CP violation since we can't ignore
the matter effect. I give a simple expression for T violation
in the presence of matter.
\end{abstract}

\end{titlepage}

\newpage
\renewcommand{\thepage}{\arabic{page}}

\section{Introduction}

The CP and T violation is a fundamental and important problem of the
particle physics and cosmology.
We have examined CP violation in the quark sector.
Is there CP violation in the lepton sector?
 
The answer is 'Yes' if the neutrinos have 
masses and complex mixing angles in the electroweak theory.

The neutrino oscillation search is a powerful experiment which can
examine masses and mixing angles of the neutrinos.
In fact the several underground experiments have shown
lack of the solar
neutrinos\cite{Ga1,Ga2,Kam,Cl}
and anomaly in the atmospheric
neutrinos\cite{AtmKam,IMB,NUSEX,SOUDAN2,Frejus},
implying that the neutrinos have masses.
The solar neutrino deficit implies the mass difference
of $10^{-5}-10^{-4}$ eV$^2$.
atmospheric neutrino anomaly suggests mass difference
around $10^{-3} \sim 10^{-2}$
eV$^2$\cite{Fogli1,FLM,Yasuda}.

The latter fact encourages us to make
long base line neutrino experiments. Recently such
experiments are planned and will be operated in the
near future\cite{KEKKam,Ferm}. It seems necessary for 
us to examine whether there is a chance to observe 
not only the neutrino oscillation but also the CP or
T violation by
long base line experiments. Here I show
such possibilities
taking account of the atmospheric neutrino experiments and
considering the solar neutrino experiments. I consider
T violation mainly.
This talk is based on the work with J. Arafune\cite{AS}.

\section{Formulation of
CP and T violation in neutrino oscillation}

\subsection{Assumptions}
First I show assumtions in this letter.

I assume there are three generations of neutrinos
$\nu_e, \nu_\mu$ and $\nu_\tau$. I do not assume the existence
of a sterile neutrino.

I also assume that the solar neutrino deficit and the atmospheric
neutrino anomaly are attributed to neutrino oscillation, that is
one of the two mass square
difference is around $10^{-5}-10^{-4}$ eV$^2$
and the other is in the range of $10^{-3}-10^{-2}$ eV$^2$.
I denote the smaller mass square difference by $\delta m_{21}^2$
and the larger by $\delta m_{31}$.
\begin{eqnarray}
\delta m_{21}^2 \sim 10^{-5}-10^{-4} {\rm eV}^2
\nonumber\\
\delta m_{31}^2 \sim 10^{-3}-10^{-2} {\rm eV}^2
\label{eq:mass}
\end{eqnarray}

\subsection{Brief review}

I briefly review CP and T
violation in vacuum oscillation\cite{yana,BP,Pakvasa}.

I will use the parametrisation for mixing matrix $U$
by Chau and Keung\cite{CK,KP,Toshev},
\begin{eqnarray}
U &=& \pmatrix{1&0&0\cr 0&c_\psi&s_\psi\cr 0&-s_\psi&c_\psi}
\pmatrix{1&0&0\cr 0&1&0\cr 0&0&e^{i \delta}}
\pmatrix{c_\phi&0&s_\phi\cr 0&1&0\cr -s_\phi&0&c_\phi}
\pmatrix{c_\omega&s_\omega&0\cr -s_\omega&c_\omega&0\cr 0&0&1}
\label{eq:mixingmatrix}
\\
&\equiv&U_\psi \Gamma U_\phi U_\omega,\nonumber
\end{eqnarray}
which is used by Particle Data Group.

The T violation gives the difference
between the transition probability of
$\nu_\alpha \rightarrow \nu_\beta$ and that of
$\nu_\beta \rightarrow \nu_\alpha$\cite{BWP}:
\begin{eqnarray}
&&P(\nu_\alpha \rightarrow \nu_\beta; E, L) -
P(\nu_\beta \rightarrow \nu_\alpha; E, L)
\nonumber \\
&=&4 \sin\omega \cos\omega \sin\psi \cos\psi
\sin\phi \cos^2\phi \sin\delta
(\sin\Delta_{21}+\sin\Delta_{32}+\sin\Delta_{13})
\label{eq:BWP}
\\
&\equiv& 4 J f,
\nonumber
\end{eqnarray}
where 
\begin{eqnarray}
\Delta_{ij}&\equiv&\delta m_{ij}^2 {L\over 2E}
= 2.54 {(\delta m_{ij}^2/10^{-2}{\rm eV}^2)
\over (E/{\rm GeV})} (L/100{\rm km}),\nonumber\\
J &\equiv& \sin\omega \cos\omega \sin\psi \cos\psi
\sin\phi \cos^2\phi \sin\delta
\nonumber\\
&&\nonumber\\
f &\equiv&
(\sin\Delta_{21}+\sin\Delta_{32}+\sin\Delta_{13}).
\nonumber
\end{eqnarray}
$\delta m_{ij}^2$ is mass square difference.
$L$ is the distance between the production point and the detection
point. $E$ is the neutrino energy.

In the vacuum the CPT theorem gives the relation between
the transition probability of
anti-neutrino and that of neutrino,
\begin{eqnarray}
P(\bar\nu_\alpha \rightarrow \bar\nu_\beta; E, L) =
P(\nu_\beta \rightarrow \nu_\alpha; E, L),
\end{eqnarray}
which relates CP violation to T violation:
\begin{eqnarray}
&&P(\nu_\alpha \rightarrow \nu_\beta; E, L) -
P(\bar\nu_\alpha \rightarrow \bar\nu_\beta; E, L)
\nonumber \\
&=&P(\nu_\alpha \rightarrow \nu_\beta; E, L) -
P(\nu_\beta \rightarrow \nu_\alpha; E, L).
\label{eq:cpt}
\end{eqnarray}

\subsection{CP and T violation in long base line experiments}

Let's consider how large the T(CP) violation
can be in long base line experiments with
the assumed mass square differences eq.\ref{eq:mass}.
In long baseline experiments neutrinos have energy of
several GeV's and the distance of several hundreds km.
$|\Delta_{31}| \sim O(1)$ and $|\Delta_{21}| =
|\epsilon \Delta_{31}| \ll 1$ in this case, where
$\epsilon \equiv {\delta m_{21} \over \delta m_{31}}$.
The oscillatory part $f$
becomes $O(\epsilon)$:
\begin{eqnarray}
f(\Delta_{31},\epsilon)
&=&\sin\Delta_{21}+\sin\Delta_{32}+\sin\Delta_{13}
\nonumber\\
&=&\sin(\epsilon\Delta_{31})+\sin\{(1-\epsilon)\Delta_{31}\}
-\sin\Delta_{31}
\label{eq:f}\\
&=& \epsilon \Delta_{31}(1-\cos\Delta_{31})
+O(\epsilon^2 \Delta_{31}^2).
\label{eq:fapp}
\end{eqnarray}

\begin{figure}
\unitlength=1cm
\begin{picture}(16,6)
\unitlength=1mm
\put(25,80){$f(\Delta_{31},\epsilon = 0.03)$}
\put(150,3){$\Delta_{31}$}
\centerline{
\epsfxsize=13cm
\epsfbox{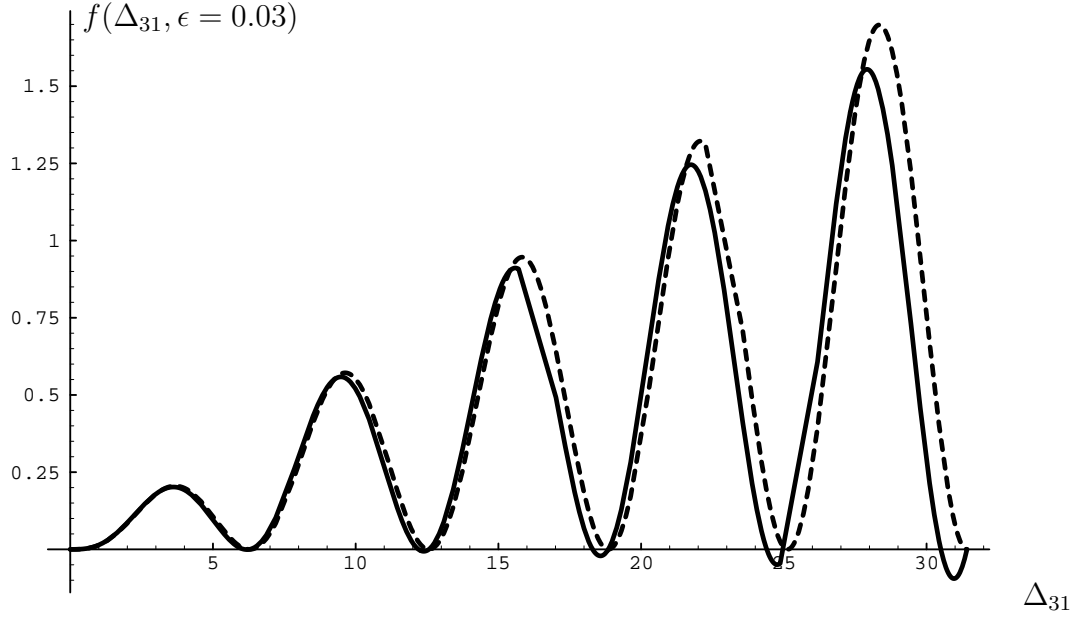}
}
\end{picture}
\caption{Graph of $f(\Delta_{31},\epsilon)$
for $\epsilon = 0.03$.
The solid line and the dashed line represent the exact expression
eq.\ref{eq:f} and the approximated one eq.\ref{eq:fapp}, respectively.
The approximated $f$ has peaks
at $\Delta_{31}=3.67, 9.63,
15.8, \cdots$ irrespectively of $\epsilon$.
}
\label{fig1}
\end{figure}

Fig.\ref{fig1} shows the graph of
$f(\Delta_{31},\epsilon=0.03)$. The approximation eq.\ref{eq:fapp}
works very well up to $|\epsilon \Delta_{31}| \sim 1$.
In the following we will use
eq.\ref{eq:fapp} instead of eq.\ref{eq:f}.
We see many peaks of $f(\Delta_{31},\epsilon)$ in fig.\ref{fig1}.
In practice, however, we do not see such sharp peaks
but observe the value averaged around there, for
$\Delta_{31}$ has a spread due to the energy spread of
neutrino beam ($|\delta\Delta_{31}/\Delta_{31}| = |\delta E/E|$).
In the following we will assume $|\delta\Delta_{31}/\Delta_{31}|
= |\delta E/E| = 20\%$\cite{Nishikawa} as a typical value.

Table \ref{table1} gives values of
$f(\Delta_{31},\epsilon)/\epsilon$ at the first several peaks
and the averaged values around there.

We see the T violation effect,
\begin{eqnarray}
<P(\nu_\alpha \rightarrow \nu_\beta)
- P(\nu_\beta \rightarrow \nu_\alpha)>_{20\%}
= 4 J <f>_{20\%} = J \epsilon \times
\left\{
\begin{array}{c}
25.9 \\
56.0 \\
62.4\\
\vdots
\end{array}
\right.
\hbox{\ for\ }
\Delta_{31} =
\left\{
\begin{array}{c}
3.67 \\
9.63 \\
15.8 \\
\vdots
\end{array}
\right.
\label{eq:atbest2}
\end{eqnarray}
at peaks for neutrino beams with 20 \% of energy spread.
Note that the averaged peak values decrease 
with the spread of neutrino energy.

Which peak we can reach depends on
$\delta m_{31}^2, L$ and $E$. The first peak
$\Delta_{31} = 3.67$ is reached, for example
by $\delta m_{31}^2 = 10^{-2}$ eV$^2$,
$L=250$ km
(for KEK-Kamiokande long base line experiment)
and neutrino energy $E = 1.73$ GeV.

\begin{table}
\begin{center}
\begin{tabular}{c|c|c|c}
$\Delta_{31}$ & \ $f/\epsilon$\  & $< f/\epsilon>_{10\%}$ &
$<f/\epsilon>_{20\%}$\\
\hline&&&\\
3.67&6.84&6.75&6.48\\
9.63&19.1&17.6&14.0\\
15.8&31.5&25.7&15.6\\
\vdots&\vdots&\vdots&\vdots
\end{tabular}
\end{center}
\caption{The peak values of
$f(\Delta_{31},\epsilon)/\epsilon$ and the corresponding averaged
values. Here
$<f/\epsilon>_{20\%(10\%)}$ is
a value of $f(\Delta_{31},\epsilon)/\epsilon
=\Delta_{31}(1-\cos\Delta_{31})$(see eq.\ref{eq:fapp}) averaged 
over the range $0.8\Delta_{31}\sim 1.2\Delta_{31}$
($0.9\Delta_{31}\sim 1.1\Delta_{31}$).
}
\label{table1}
\end{table}

\vspace*{0.5cm}

\section{T violation}

Neutrinos, however, go through not vacuum but matter, the surface
of the earth. There is an effect by matter. The matter effect is
calculated like this\cite{Wolf,MS}:
\begin{eqnarray}
&&2\sqrt{2}G_F n_e E \sim 2\times10^{-4} {\rm eV}^2
\left({E \over {\rm GeV}}\right)
\left({n\over 3{\rm g/c.c}}\right),
\label{eq:mattereffect}
\end{eqnarray}
where $n_e$ is the electron number density of the
earth and $n$ is the matter density of the surface of the earth.
For neutrino with several GeV energy the matter
effect is much greater than the smaller mass square difference
$\delta m^2_{21}$, though it will be much smaller than
the larger mass square difference.
Thus we cannot see pure CP violating effect.
It requires to subtract such effect in order to
deduce the pure CP violation effect\cite{prep}.
In principle it is possible,
because the matter effect is proportional to $E$
while $\delta m^2_{21}$ is constant.

In the matter with  constant density\footnote{
Note that the time reversal of $\nu_\alpha \rightarrow
\nu_\beta$ requires the exchange of the production point and the
detection point and the time reversal of
$P(\nu_\alpha \rightarrow \nu_\beta)$
in matter is in general different from
$P(\nu_\beta \rightarrow\nu_\alpha)$\cite{KP}.},
 we have a pure
T violation effect $P(\nu_\alpha \rightarrow
\nu_\beta) - P(\nu_\beta \rightarrow
\nu_\alpha)$, though we do not observe a pure CP violation effect
because of an apparent CP violation due to matter.
The pure T violation effect is given by the eq.\ref{eq:BWP}
with mixing matrix and mass square differences in matter.

\subsection{T violation in matter}

To calculate how large the T violation is, we have to know
the mixing matrix and mass square differences in matter.
When a neutrino is
in matter, its matrix of effective mass squared
$M_m^2$ for weak eigenstates is\cite{KP,Toshev}
\begin{eqnarray}
M_m^2 = U \pmatrix{0&&\cr&\delta m_{21}^2&\cr&&\delta m_{31}^2}
U^\dagger + \pmatrix{a&&\cr&0&\cr&&0},
\end{eqnarray}
where $a = 2\sqrt{2} G_F n_e E$ and $U$ is given by
eq.\ref{eq:mixingmatrix}.
This is diagonalized by a
mixing matrix $U_m$ as $M_m^2 = U_m
{\rm diag}(\tilde m_1^2,\tilde m_2^2, \tilde m_3^2)
U_m^\dagger$. 

Up to the leading order of $\delta m_{21}^2$, the
the eigenvalues of effective mass square($\Lambda_i$) and the mixing
matrix $U_m$ are given by,
\begin{eqnarray}
\Lambda_1 &=& {(a+\delta m^2_{31}) - \sqrt{(a+\delta m^2_{31})^2-
4a\delta m^2_{31}\cos^2\phi} \over 2},\\
\Lambda_2 &=& 0,\\
\Lambda_3 &=& {(a+\delta m^2_{31}) + \sqrt{(a+\delta m^2_{31})^2-
4a\delta m^2_{31}\cos^2\phi} \over 2},
\end{eqnarray}
and
\begin{eqnarray}
U_m=U_\psi \Gamma U_{\phi'} \tilde U
\nonumber
\end{eqnarray}
where
\begin{eqnarray}
\tan 2\phi'&=& {\delta m^2_{31} \sin 2\phi \over
\delta m^2_{31}\cos 2\phi-a},
\nonumber\\
\tilde U &=& \delta_{ij}+\tilde U', \tilde U'^T=-\tilde U',
\nonumber
\end{eqnarray}
with
\begin{eqnarray}
(\tilde U')_{12}&=&{\delta m^2_{21} \over \Lambda_1} \cos (\phi-\phi')
\cos\omega \sin\omega
\nonumber\\
(\tilde U')_{13}&=&{\delta m^2_{21} \over \Lambda_3-\Lambda_1} 
\cos (\phi-\phi') \sin (\phi-\phi') \sin^2\omega
\nonumber\\
(\tilde U')_{23}&=&{\delta m^2_{21} \over \Lambda_3} \sin (\phi-\phi')
\cos\omega \sin\omega.
\nonumber
\end{eqnarray}

Using them, $J$ in matter, $J_m$, takes the form.
\begin{eqnarray}
J_m 
= - {\delta m_{21}^2 \over a} 
{\delta m_{31}^2
\over \{(\delta m_{31}^2 + a)^2 - 4\delta m_{31}^2 a
\cos^2\phi\}^{1/2}}
\sin\omega\cos\omega\sin\psi\cos\psi\sin\phi\sin\delta.
\label{eq:Jmp}
\end{eqnarray}

\subsection{Most likely case:
$\delta m^2_{21} \ll a \ll \delta m^2_{31}$}

For example we consider this case because as mentioned
earier this case seems most likely to be realized.
I show the result calculated up to the leading order
of two small parameter $\epsilon_1 \equiv {a\over\delta m_{31}^2}$
and $\epsilon_2 \equiv {\delta m_{21}^2\over a}$.

Then we have the effective masses
\begin{eqnarray}
\tilde m^2_1 &\simeq& \Lambda_1 \simeq a\cos^2\phi ,
\nonumber\\
\tilde m^2_2&\simeq&  \Lambda_2 \simeq 0 ,\\
\tilde m^2_3&\simeq&  \Lambda_3 \simeq 
\delta m_{31}^2 + a\sin^2\phi.
\nonumber
\end{eqnarray}
and ``mass square difference ratio'' in matter
\begin{eqnarray}
\epsilon_m = 
{\tilde m^2_2 - \tilde m^2_1 \over \tilde m^2_3 - \tilde m^2_2}
\simeq - { a\cos^2\phi \over \delta m_{31}^2}.
\end{eqnarray}
Note that $|\epsilon_m| \ll 1$.

$J_m$ is given by,
\begin{eqnarray}
J_m \sim  - {\delta m_{21}^2 \over a}
\sin\omega\cos\omega\sin\psi\cos\psi\sin\phi\sin\delta.
\nonumber
\end{eqnarray}

The key quantity $J\epsilon$ in matter(see eq.\ref{eq:atbest2}) is then,
\begin{eqnarray}
J_m \epsilon_m = J \epsilon,
\end{eqnarray}
same as that of vacuum.

Using the argument similar to that used to derive
equation for T violation in vacuum,
we obtain the T violation effect
\begin{eqnarray}
<P(\nu_\alpha \rightarrow \nu_\beta)
- P(\nu_\beta \rightarrow \nu_\alpha)>_{20\%}
= J_m \epsilon_m \times
\left\{
\begin{array}{c}
25.9 \\
56.0 \\
62.4\\
\vdots
\end{array}
\right.
= J \epsilon \times
\left\{
\begin{array}{c}
25.9 \\
56.0 \\
62.4\\
\vdots
\end{array}
\right.,
\end{eqnarray}
at peaks, where
we choose the mean neutrino energy $E$ to satisfy
(see Table \ref{table1})
\begin{eqnarray}
\Delta_{31} = \delta m^2_{31} {L \over 2E} = 3.67, 9.63, 15.8
\ldots
\end{eqnarray}

The size of T violation in matter is same as that in vacuum.
Incidentally I may remark that there is no correction of 
O($\epsilon_2$) to this value because the limit $a \rightarrow 0$
is smooth.

In the case of large mixing angles
\cite{FLM}, $J/\sin\delta$
$\sim0.06$ and
 $\epsilon \sim 10^{-2}$ are allowed\footnote{
Here $\sin\omega \sim 1/2, \sin\psi\sim1/\sqrt{2}$
and $\sin\phi
= \sqrt{0.1}$.} 
 for example.
Then the size of T violation
\begin{eqnarray}
<P(\nu_\alpha \rightarrow \nu_\beta) - P(\nu_\beta \rightarrow
\nu_\alpha)>_{20\%}
= \left({J/\sin\delta\over 0.06}\right)
\left({\epsilon\over 10^{-2}}\right) \sin\delta \times
\left\{
\begin{array}{c}
0.015 \\
0.033 \\
0.037\\
\vdots
\end{array}
\right. .
\end{eqnarray}
We may see a few \% of T violation.

Numerically it is also confirmed with slowly changing density,
which satisfies the condition $a \ll \delta m^2_{31}$,
up to O($\epsilon_1$). I assumed the density takes the following
form:
\begin{eqnarray}
a=a_0+a_1\cos k_1(x-x_1) +a_2\cos k_2(x-x_2),
\nonumber
\end{eqnarray}
with $|a_0+a_1+a_1| \ll \delta m^2_{31}$
and various $a_i,k_i$ and $x_i$.

\section{Summary}

We have examined mainly T violation in the long base
line experiment with the assumption
that $\delta m^2_{21}$
is much smaller than the matter effect 
``$a$'' and $\delta m^2_{31}$,
that is both the solar neutrino deficit and atmospheric
neutrino anomalies are attributed to
the neutrino oscillation
violation effect. In this case we cannot see pure
CP violation because of matter.
A fer \% of T violation is, however, observable depending on
the parameters for neutrino. More exactly we may see
T violation of O(${\delta m_{21}^2\over\delta m_{31}^2}$).

The reason of  O(${\delta m_{21}^2\over\delta m_{31}^2}$)
is as follows:
The probability of CP and T violation
effect
should vanish for  $\delta m^2_{21} \rightarrow$ 0, 
and therefore be proportional to  
$\delta m^2_{21}/\delta m^2_{31}$, $\delta m^2_{21}/(E/L)$
or $\delta m^2_{21}/a$ by the dimensional analysis. We expected
an enhancement for T violation, that is, it is proportional
to not O(${\delta m_{21}^2\over\delta m_{31}^2}$)
but O(${\delta m_{21}^2\over a}$). We have, however,
CP and T violation of O(${\delta m_{21}^2\over\delta m_{31}^2}$)
because in the limit that $a \rightarrow 0$
they must be smoothly connected to
the vacuum case.


\begin{thebibliography}{99}
\bibitem{Ga1} GALLEX Collaboration, P. Anselmann {\it et al},
Phys. Lett. {\bf B357},(1995) 237.
\bibitem{Ga2} SAGE Collaboration, J. N. Abdurashitov {\it et al},
Phys. Lett. {\bf B328}, (1994) 234.
\bibitem{Kam} Kamiokande Collaboration, Y. Suzuki, Nucl, Phys.
Proc. Suppl. {\bf B38}, (1995) 54.
\bibitem{Cl} Homestake Collaboration, B. T. Cleveland {\it et al},
Nucl, Phys. Proc. Suppl. {\bf B38}, (1995) 47.
\bibitem{AtmKam} Kamiokande Collaboration,
K.S. Hirata et al., Phys. Lett. {\bf B205} (1988) 416;
{\it ibid.} {\bf B280} (1992) 146;
 Y. Fukuda et al., Phys. Lett. {\bf 335B}(1994)237.
\bibitem{IMB} IMB  Collaboration,
D. Casper et al., Phys. Rev. Lett. {\bf 66} (1991) 2561;\\
R. Becker-Szendy et al., Phys. Rev. {\bf D46} (1992) 3720.
\bibitem{NUSEX}  NUSEX Collaboration, M. Aglietta  {\it et al},
Europhys. Lett. {\bf 8}(1989)611;
{\it  ibid} {\bf 15}(1991)559.
\bibitem{SOUDAN2} SOUDAN2 Collaboration, T. Kafka,
Nucl. Phys.Proc. Suppl. {\bf B35} (1994) 427; M. Goodman
{\it ibid} {\bf 38} (1995) 337.
\bibitem{Frejus} Fr\'ejus Collaboration, K. Daum {\it et al},
Z. Phys. {\bf C66}(1995)417.
\bibitem{Fogli1}  G. L. Fogli, E. Lisi, D. Montanino
and G.Scioscia, IASSNS-AST 96/41
(hep-ph/9607251).
\bibitem{FLM} G. L. Fogli, E. Lisi and D. Montanino, Phys. Rev
{\bf D49}, 3626 (1994).
\bibitem{Yasuda} O. Yasuda, TMUP-HEL-9603
(hep-ph/9602342).
\bibitem{KEKKam} K. Nishikawa, INS-Rep-924(1992).
\bibitem{Ferm}
S. Parke, Fermilab-Conf-93/056-T(hep-ph/9304271)(1993).
\bibitem{AS} J. Arafune and J. Sato, ICRR-Report-369-96-20
(hep-ph/9607437), to appear in Phys. Rev. D.
\bibitem{yana} For a review, M. Fukugita and T. Yanagida,
 in {\it Physics and Astrophysics of Neutrinos}, edited by
 M. Fukugita
and A. Suzuki, (Springer-Verlag, Tokyo, 1994).
\bibitem{BP} S. M. Bilenky and S. T. Petcov, Rev. Mod. Phys.
{\bf 59}, 671 (1987).
\bibitem{Pakvasa} S. Pakvasa, in High Energy Physics-1980
(20th Int. Conf. on High Energy Physics),
ed.L.Durand and L.Pondrom, vol.2,p.1164.
\bibitem{CK} L.-L. Chau and W.-Y. Keung, Phys. Rev. Lett. {\bf 53},
1802 (1984).
\bibitem{KP} T.K Kuo and J. Pantaleone, Phys. Lett. {\bf B198},
406 (1987).
\bibitem{Toshev} S. Toshev, Phys. Lett. {\bf B226}, 335 (1989).
\bibitem{BWP} V. Barger, K. Whisnant and R. J. N. Phillips,
Phys. Rev. Lett. {\bf 45}, 2084 (1980).
\bibitem{Nishikawa} Nishikawa, private communication.
\bibitem{Wolf} L. Wolfenstein, Phys. Rev. {\bf D17}, 2369 (1978).
\bibitem{MS}
S. P. Mikheev and A. Yu. Smirnov, Sov. J. Nucl. Phys. {\bf 42},
 913 (1985)
\bibitem{prep}J.~Arafune, M.~Koike and J.~Sato, in preparation.
\end{thebibliography}
\end{document}